\documentclass[reprint,
superscriptaddress,
amsmath,amssymb,aps,showkeys,showpacs,
twoside,final,secnumarabic,
nofootinbib]{revtex4-2}

\usepackage[paperwidth=205mm,paperheight=290mm,top=17mm,bottom=25mm,
inner=17mm,outer=17mm,
twoside]{geometry}

\usepackage{cmap} 
\usepackage[T1,T2A]{fontenc}
\usepackage[utf8]{inputenc}
\usepackage[russian,english]{babel}
\usepackage{color}
\usepackage{graphicx}
\usepackage{dcolumn}
\usepackage{bm} 
\usepackage[unicode=true,colorlinks=true,linkcolor=magenta, urlcolor=blue, citecolor = blue,breaklinks]{hyperref}
\usepackage{multirow}
\usepackage{url}
\usepackage{placeins}
\usepackage{breakurl}
\DeclareGraphicsExtensions{.eps}

\newcount\issue
\newcount\Vol
\newcount\numb
\usepackage{fancyhdr} 
\def\Vol{\textbf{79}}
\def\numb{x}
\begin{document}

\title{Machine Learning for Environmental Sciences
 \\[20pt] Probabilistic programming methods for reconstruction \\of multichannel imaging detector events: ELVES and TRACKS} 

\def\addressa{Skobeltsyn Institute of Nuclear Physics, Lomonosov Moscow State University, Moscow 119991, Russia}
\def\addressb{Faculty of Physics, Lomonosov Moscow State University, Moscow 119991, Russia}

\author{\firstname{S.A.}~\surname{Sharakin}}
\email[E-mail: ]{sharakin@mail.ru}
\affiliation{\addressa}
\author{\firstname{R.E.}~\surname{Saraev}}
\affiliation{\addressa}
\affiliation{\addressb}

\received{xx.10.2024}
\revised{xx.xx.2024}
\accepted{xx.xx.2024}

\begin{abstract}
This paper proposes new methods for analyzing dynamic images registered by multichannel, highly sensitive detectors with \textit{low spatial but high temporal resolution}. The principal characteristic of the approach is the absence of factorization of different types of information within the data set. For a number of rapidly changing (transient) phenomena in the Earth's atmosphere, a probabilistic model can be formulated, and the parameters of this model can be reconstructed using \textit{probabilistic programming} methods (Bayesian inference based on Markov chain Monte Carlo). This paper demonstrates the aforementioned approach on a number of examples, both simulated and actually registered by the detectors of the SINP MSU.

In the case of submillisecond ELVES events registered by the orbital Mini-EUSO detector on board the ISS, the probabilistic model includes the coordinates and orientation of the lightning discharge that generated the glow, as well as the height of the ionized layer in which the glow is registered, among its parameters. Bayesian inference, implemented by means of the PyMC library, allows us to calculate posterior distributions for these parameters based on the times of signal peaks in individual detector channels.

In addition to studying different types of aurora, the circumpolar system of ground-based multichannel PAIPS detectors also serves as a test-bench for probabilistic reconstruction algorithms. A wide class of track events is used for this purpose - meteors, satellite and aircraft passes, and the movement of stars across the sky. The Bayesian model includes both the parameters of the track event itself and the peculiarities of its registration. These methods can be generalized to stereo events (track registration by two detectors with overlapping fields of view) or applied to the reconstruction of extremely high energy cosmic rays in orbital fluorescence detectors.
\end{abstract}

\keywords{probabilistic programming, Bayesian inference, transient events, imaging detectors   \\[5pt]}

\maketitle
\thispagestyle{fancy}


\section{Introduction}\label{intro}

We have become so accustomed to high-quality photographs obtained both from the Earth's surface and from space that we sometimes begin to forget that there are areas of experimental science in which it will not be possible in principle to obtain such a high-precision image in the near future. One of these areas is the registration of transient phenomena in the Earth's atmosphere, in natural - not controlled, not laboratory - conditions. Transient means rapidly variable, and therefore the recording device, the detector of transients, should be very fast (with high temporal resolution) and highly sensitive. Therefore, transient image detectors often use photomultiplier tube (PMT) matrices with relatively large pixel sizes instead of CCDs. As a consequence, such devices have low spatial (angular) resolution, and the data recorded are a set of coarse-grained photographs.

However, these detectors provide us with invaluable information on the spatial and temporal dynamics of a variety of fast processes in the atmosphere, from meteors to ultra high energy cosmic rays (UHECR)~\cite{Coleman23_UHECR}.

Transient detectors have been developed at the Skobeltsyn Institute of Nuclear Physics of Moscow State University for the last two decades, mainly for the registration of optical signals from low-Earth orbit. These detectors are based on a multichannel photoreceiver that registers radiation in the near-UV (300-400~nm). The first such device was TUS~detector on the Lomonosov satellite \cite{Barghini2022_TUSresults}, which was operational in orbit between~2016 and~2017. It comprised 256~channels ('pixels') with a field of view of 5~by 5~km at the surface, offering a temporal resolution of 800~ns. At present, the UV~Atmosphere (Mini-EUSO\footnote{Multiwavelength Imaging New Instrument for the Extreme Universe Space Observatory.}) detector has been operating on the~ISS for more than four years already \cite{Adams2015_MiniEUSOScience}. It was developed within the framework of the international collaboration JEM-EUSO\footnote{Joint Exploratory Missions for an Extreme Universe Space Observatory. The Russian side in the collaboration is represented by SINP MSU.}, and consists of a matrix of~2304 channels and a system of Fresnel lenses, providing focusing in the field of view of 44 by 44 degrees. The temporal resolution of the instrument is 2.5~$\mu$s. 
There are a number of JEM-EUSO balloon and ground experiments \cite{Abdellaoui2018_EUSO-TA}, \cite{Adams2022_EUSO-Ball}, \cite{Abdellaoui2024_EUSO-SPB1} and several more space missions are being prepared \cite{Klimov2022_KEUSO}, \cite{Olinto2021_POEMMA}.
A characteristic feature of the JEM-EUSO line of detectors is the modular structure of the photoreceiver with the presence of rather large ``dead'' zones between its individual modules (``elementary cells'') and individual PMTs.

Both telescopes, TUS and Mini-EUSO, have recorded a large amount of ``coarse-grained'' video with physically unique content \cite{Marcelli2024_MiniEUSOresults}. This paper will focus on two types of transient events, elves and meteors, and how to extract information from these data using probabilistic models and Bayesian inference \cite{Sivia2006_BI}, \cite{Martin2018_BI}.

\section{ELVES Reconstruction}
\label{sec:ELVES_reco}
In thunderstorm conditions, along with processes such as lightning discharges, \textit{transient} luminous atmospheric phenomena can be registered from low-Earth orbit. These include sprites and their halos, elves, blue starters and jets, giant jets, and some other \cite{Newsome2010_TLE}, \cite{Pasko2012_TLE}. Some of the fastest of these are ELVES (Emission of Light and Very low frequency perturbations due to Electromagnetic pulse Sources), which manifest as sub-millisecond glows at altitudes $\sim90$~km in the form of a rapidly expanding ring \cite{Fukunishi1996_ELVES}. This glow is caused by the interaction of the electromagnetic pulse generated by lightning (or other preceding discharge processes in a thundercloud) with the lower layer of the ionosphere. A spherical wavefront crosses the relatively thin layer of the ionosphere and makes it glow \cite{Barrington-Leigh1999_ELVES}. This phenomenon is of truly gigantic scale, with the diameter of the luminous ring extending for hundreds of kilometres. Highly sensitive orbital detectors have registered ELVES with diameters reaching 1000~km \cite{Klimov2019_TUSelves}, \cite{Sharakin2024_RecoELVES}.

ELVES is an invaluable source of information regarding the processes occurring within a thunderstorm cloud, which can be considered a unique ionospheric ``fingerprint'' of thunderstorm activity. The conventional method for recovering information from these fingerprints entails processing each frame sequentially. This involves approximating the instantaneous image by a ring with its centre and outer or central radius, followed by calculating the average position of the centre and subsequently recalculating the kinematics of the ring's development. Only after this procedure is complete can the localization of the discharge causing the event be estimated.

In this approach, the geometric information in the data is pre-separated from the temporal information. Due to the low spatial resolution of the orbital detector, this factorization appears to be particularly inefficient, especially in the case of large radius rings and significant and inhomogeneous background illumination over the field of view.

It seems reasonable to propose a model of ELVES reconstruction that would be initially \textit{kinematic}. In this case, the entire set of information regarding the development of luminescence in space and time – the \textit{spatiotemporal pattern} represented by the times of the signal peak in the field of view of each detector channel can be considered as data, Data = $\{ T_\mathrm{peak}[i], \theta[i], \phi[i] \}$. Here $i$ is the identifier of the channel with the signal from the ELVES (active signal - with pre-subtracted background), the angles $\theta$ and $\phi$ set the positions of the field of view of this channel (the centre of the ``pixel''). The $T_\mathrm{peak}$ is calculated by approximating the active signal in the channel by an asymmetric parametric profile that reflect the rapid growth of the signal (the arrival of the electromagnetic wave front in the ionospheric luminescence zone).

It is convenient to take into account different types of information by means of a probabilistic model that includes both parameters of the phenomenon itself (in this case the localization of the discharge and the height of the ionospheric layer) and characteristics of the measuring equipment - errors in the determination of times and fields of view, characteristics of the detector response function (in particular the shape and size of the point spread function (PSF) of the wide-angle optics). This approach was first used for the ELVES reconstruction in \cite{Sharakin2024_RecoELVES}.

To construct the parametric model of the~ELVES, we introduce a Cartesian coordinate system with the centre~O on the ground surface below the current position of the~ISS (at the ``nadir'' of the~ISS) and the vertical $z$-axis (the $x-y$ plane is horizontal). In this coordinate system, the position of the ``discharge point'' S$(x_0, y_0, z_0)$, the ``emission point'' E$(x, y, z)$ and the detector D$(0, 0, H_\mathrm{d})$ are given, which allows us to represent the (modelled) time required for the EM~wave to propagate from the discharge point to the emission point and for the subsequent propagation of the radiation to the detector as follows
\begin{eqnarray}
\label{eq:DTmod}
&& c\Delta T_\mathrm{mod}(x,y) = \sqrt{x^2+y^2+(H_\mathrm{d} - H_\mathrm{e})^2} \nonumber \\
&& + \sqrt{(x-x_0)^2+(y-y_0)^2+(H_\mathrm{e} - z_0)^2}
\end{eqnarray}
where $c\approx 0.3$~km/$\mu$s is the speed of light, and the orbital height $H_\mathrm{d}\simeq 420$~km is assumed to be known (as part of the the telemetry information). In~(\ref{eq:DTmod}) it is taken into account that the $z$-coordinate of the emission point~E can be expressed through one more parameter of the model - the height~$H_\mathrm{e}$ of the ionospheric layer on which the glow is observed: estimates show that inside the detector's field of view the difference between~$z$ and~$H_\mathrm{e}$ can be neglected. From the literature data $H_\mathrm{e} \approx 90$~km \cite{Inan1991_Ionosph}, but the uncertainty of this value can be several kilometres \cite{Fukunishi1996_ELVES}.

Similarly, $z_0$ is expressed in terms of the (local) discharge height~$H_0$. However, for individual events, the radial distance $\rho_0 = \sqrt{ x_0^2+y_0^2}$ to the discharge can be hundreds of kilometres (Mini-EUSO registered ELVES with $\rho_0$ up to 500~km), so it is necessary to take into account a correction for the ``sphericity'' of the atmosphere: $z_0 \approx H_0 - \rho_0^2/(2R_\mathrm{E})$, where $R_\mathrm{E} = 6380$~km is the average radius of the~Earth.

Thus, the kinematic model of the ELVES can be described using a set of parameters $\Theta = \{ x_0, y_0, H_0, H_\mathrm{e} \}$.

Reconstruction in the framework of the probabilistic model implies the calculation (Bayesian inference), based on the data obtained in the experiment, of the posterior distribution $p(\Theta|\mathrm{Data})$ as the product (up to the normalization factor) of the likelihood function $p(\mathrm{Data}|\Theta)$, which characterises the relationship between the phenomenon and the measurement procedure, and the a~priori distribution (‘prior’) $p(\Theta)$, which allows the introduction of information additional to the experimental data into the reconstruction \cite{Martin2018_BI}.

The unknown parameters of the model in this approach cease to be ordinary numbers - they become distributions, or in the language of probabilistic programming methods - \textit{stochastic} quantities. Their stochastic nature implies setting a prior distribution $p(x_0, y_0, H_0, H_\mathrm{e})$, which is often sufficient to choose in a fully or partially factorized form, $p(\Theta) = p(x_0, y_0) p(H_0) p(H_\mathrm{e})$.

The priors can be both low-informative and act as regularizer, and can also reflect (as a probabilistic measure) the available additional information. For example, for $p(H_0)$ and $p(H_\mathrm{e})$, there are reasons to localize the discharge and glow near heights of 5 and 90~km, respectively, by setting Gaussian distributions with a fixed standard deviation (e.g., 5~km).

The likelihood actually plays the role of a measurement error model and in the case of ELVES can be expressed as
$$
T_\mathrm{peak}[i] = T_0 + \Delta T_\mathrm{mod}(x[i], y[i]; \Theta) + \xi[i], 
$$
where $T_0$ is the (unknown) discharge moment, and $\xi[i]$ are independently distributed quantities with zero mean (here we have explicitly specified the dependence of $\Delta T_\mathrm{mod}$ on the model parameters). In the simplest case of Gaussian measurement errors with equal variance\footnote{We use the notation ${\cal N}(\mu,\sigma)$ for a univariate normal distribution with mean~$\mu$ and standard deviation~$\sigma$. }
\begin{eqnarray}
\label{eq:ELVES_BM}
&& p(\mathrm{Data}| \Theta ) = \nonumber \\
&&\prod_i {\cal N}( T_\mathrm{peak}[i] - T_0 - \Delta T_\mathrm{mod}( x[i], y[i]; \Theta), \sigma_T )
\end{eqnarray}
The standard deviation~$\sigma_T$ reflects the errors associated with the determination of $T_\mathrm{peak}$: primarily the time resolution ($\sim~1$ clock cycle) and the errors of approximation by the model profile. In addition, the substitution used in~(\ref{eq:ELVES_BM})
\begin{eqnarray}
x[i] &=& (H_\mathrm{d}-H_\mathrm{e}) \tan\theta[i] \cos\phi[i], \nonumber\\
y[i] &=& (H_\mathrm{d}-H_\mathrm{e})\tan\theta[i]\sin\phi[i]\nonumber
\end{eqnarray}
implies that the signal peak is at the centre of the channel field of view, which is justified in the case of a (quasi)symmetric~PSF. However, the number of active channels also includes a large number of edge channels, i.e., those at the border of a single~PMT (including the edge of the photodetector elementary cell). In this case, there may be a shift of the peak away from the centre of the field of view.  This can be accounted for by considering a more complex model that takes into account the shape of the instantaneous image (e.g. by parameterising the~PSF). Alternatively, a simpler approach can be taken by specifying a non-Gaussian error distribution~$\xi[i]$. In \cite{Sharakin2024_RecoELVES}, we used a two-parameter Student distribution whose additional parameter, the number of degrees of freedom~$\nu$, controls the degree of ``non-Gaussianity'' of the data.

Thus, in addition to the parameters $\Theta$ of interest from the physical point of view, the probabilistic model also includes a set of auxiliary \textit{nuisance} parameters $\eta = \{T_0, \sigma_T, \nu\}$. The Bayesian approach involves setting priors on the nuisance parameters as well, followed by \textit{marginalization} on them:
$$
p( \Theta | \mathrm{Data} ) \propto \int p(\mathrm{Data}|\Theta, \eta ) p(\Theta, \eta) d\eta
$$
Marginalization, which is an important feature of the Bayesian reconstruction approach, leads to more reliable estimates than traditional nuisance parameter optimization \cite{Loredo2024_BayesIntro}. Marginalization is also useful for individual parameters from~$\Theta$: if the experiment is set up correctly and a lot of data about the phenomenon is collected, the posterior distributions are well concentrated and their univariate marginals have a Gaussian-like shape, and hence still allow us to give answers in the form of point estimates and their errors. Bivariate marginal distributions are useful for detecting correlations in the posterior estimates.

Bayesian methods have been known for a long time \cite{Loredo2013_BAstrostatistics}, but they became particularly attractive after their implementation as the high-level libraries such as PyMC, STAN, JAGS\footnote{\url{https://www.pymc.io/}, \url{https://mc-stan.org/}, \url{https://mcmc-jags.sourceforge.io/}} and others, as well as libraries that allow the analysis of stochastic variables (Arviz, preliZ\footnote{\url{https://python.arviz.org}, \url{https://preliz.readthedocs.io}}). These tools allow sampling from a posterior distribution (by specifying priors and likelihood) using Monte Carlo Markov chains, checking the samples obtained (convergence check, prior and posterior predictive checks, etc.) and analysing the results obtained. This approach is known in the literature as \textit{probabilistic programming}.

In our study we created a special application with a graphical interface that allows us to construct different probabilistic models, perform data pre-processing (pattern recognition), assign different error models, select priors and their (hyper)parameters - and again graphically analyse the obtained results. The PyMC-5 library (with NUTS sampler) was used as the main probabilistic programming engine.

A typical Mini-EUSO ELVES gives a set of 400-500 data points (time, position), which allows us to reliably establish the localization of the discharge when an informative distribution is chosen as a prior for the $H_\mathrm{e}$ based on the data known from the literature. However, in this case, the posterior distributions of luminescence height and discharge height will be strongly correlated: the information contained in the peak times is sensitive only to the difference of these heights $h_0 = H_\mathrm{e} - H_0$ (see Fig.~2 in \cite{Sharakin2024_RecoELVES}). This dependence can be decoupled if we take as additional information the times of the second peak, very often appearing in a signal. Additional peaks are related to the formation of the second ring of ELVES (see \cite{TUS-ELVES-1}, \cite{MiniEUSO-ELVES}), obtained as a result of the reflection of the electromagnetic pulse from the Earth's surface (in this case it will be sufficient to replace $H_0$ by $-H_0$ in the formulas).

Bayesian inference, implemented through probabilistic programming methods, thus allows us to formulate the ELVES reconstruction without factorizing the geometric part of the information from the kinematic one. We can go further and add dynamic information, e.g., the amplitude of the signal. In the measured data, this amplitude is not homogeneously distributed over the ring, which can be explained by the tilt of the electromagnetic discharge dipole \cite{Marshall2012_DoubleELVES}. So, with this approach, it is possible to recover the orientation of the discharge!

\section{Reconstruction of track events}
\label{sec:RecoTRACKS}

The main objective of the JEM-EUSO collaboration is to build an orbiting detector of UHECR. When a UHECR particle enters the Earth's atmosphere, a extensive atmospheric shower (EAS) of tens or even hundreds of billions of electrons is formed, causing a fluorescent glow in the air (mainly in the near-UV). From the point of view of the orbital image detector, the EAS glow is a ``quasi-meteor'' travelling at the speed of light, with a well-defined peak in light intensity (maximum of the shower development). On the focal surface (FS) of the photodetector we observe such an event as a track.

UHECR particles are extremely rare, so that their search from low-Earth orbit is a matter of the future, even if not distant. We are now working on both the hardware components of such a complex detector and the algorithms for triggering, pattern recognition and reconstruction of~EAS events. Any natural or anthropogenic phenomena can be used as test events, forming a \textit{track event} - an image localized in a small number of pixels, moving in time at a quasi-constant speed along a rectilinear direction. The set of instantaneous image centres, ``track points'', is called a track here. A typical representative of the family of track events are meteors.

The Mini-EUSO detector has recorded tens of thousands of meteors from ISS orbit. A recent paper by the  JEM-EUSO collaboration \cite{Barghini2024_MiniEUSOMET} has performed a detailed statistical analysis of 24,000 meteor events. Again, we face the fact that high-precision track reconstruction in our ``imperfect'' detector is not straightforward. Classical meteor reconstruction algorithms (the intersecting planes method \cite{Ceplecha1987_MetRecoIP}, the line of sight method \cite{Borovicka1990_MetRecoLoS}, the multi-parameter method \cite{Gural2012_MetRecoMP}, see also \cite{Vida2020_MetRecoMC})  rely on the high angular resolution of specialized meteor cameras, which are ineffective for our ``imprecise'' data. Especially if the track passes near the edge of the field of view of an individual PMT or crosses the FS dead zone (gaps between individual PMTs).

Nevertheless, a Bayesian approach may be employed again, whereby all available data, encompassing both kinematic and dynamic information, can be integrated into a unified probabilistic model:
$$
p(\Theta_\mathrm{kin}, \Theta_\mathrm{dyn}| \mathrm{Data}) \propto p(\mathrm{Data}|\Theta_\mathrm{kin}, \Theta_\mathrm{dyn}) p(\Theta_\mathrm{kin}, \Theta_\mathrm{dyn})
$$
The kinematic parameters $\Theta_\mathrm{kin} = \{ X_0, Y_0, \Phi, U, A \}$ may include the localization $(X_0, Y_0)$ of any of the track points (at a given time $k_0$), as well as the direction ($\Phi$) and speed ($U$) of movement along the track. For particularly long tracks, it is necessary to take into account the possible non-uniformity of the track motion. 
The speed of the track point changes due to the varying distance from the object to the detector and the deceleration of the object itself (in the case of meteors recorded from the ground, these two effects partially compensate each other). To a first approximation, the non-uniformity can be expressed by the acceleration parameter~$A$.

The dynamic parameters $\Theta_\mathrm{dyn}$ should allow us to describe the time dependence of the luminosity intensity (\textit{light curve}, LC). The light curves of meteors are very diverse \cite{Pecina2009_MeteorsLC}, which is partly due to the different type of fragmentation during the meteoroid ablation in the atmosphere. Excessive detail is not necessary: it is sufficient to provide the model with a rough correlation between the kinematic and dynamical components of the information. Therefore, we will limit ourselves to the simplest low-parameter LC~profiles, reflecting only a characteristic increase in luminosity intensity followed by a decrease. Such profiles can, for example, be composed of linear~(LIN), exponential~(EXP) and Gaussian~(GAUS) sections with characteristic signal rise and fall times~$\tau_r$, $\tau_d$. Thus, in the simplest case of a single peak track event with signal amplitude $I_\mathrm{peak}$, we have $\Theta_\mathrm{dyn} = \{ I_\mathrm{peak}, \tau_r, \tau_d \}$.

In \cite{Barghini2024_MiniEUSOMET}, the dynamic information was used only to estimate the instantaneous position of the track point as the \textit{barycentre} of the active signals, i.e. the sum of the coordinates of the centres of the corresponding pixels weighted by the instantaneous values of the signals. This approach has a number of drawbacks that lead to significant systematic errors in the reconstruction of track speed and direction. First, at weak signal (at the ``tails'' of the light curve), the sensitivity of the barycentre to the displacement of track points is greatly reduced: the $X$- or $Y$-coordinate of the barycentre (or both) stops changing, leading to underestimation of the reconstructed velocity and changes in its direction (see, for example, Fig.~4 in \cite{Barghini2024_MiniEUSOMET}). 
Secondly, the movement of the barycentre along the~FS becomes very uneven when approaching the edge of the~PMT (part of the signal disappears in the dead zone). These and a number of other shortcomings can be reduced by a more accurate determination of the instantaneous position of the track point (e.g. taking into account the shape and size of the~undersampled PSF \cite{Lauer1999_PSF} and the location of the~FS dead zones). However, in the Bayesian approach presented above, this is done automatically if the instantaneous values of the active signals in all triggered photodetector channels are selected as data, $\mathrm{Data} = \{ S_k[i] \}$, where again $i$ is the channel ID, $k$ is the clock number.

A number of auxiliary parameters will have to be introduced to form the measurement model. In the simplest case of independent Gaussian errors with the same standard deviation~$\sigma_S$
\begin{eqnarray}
&&p( S_k[i] | \Theta_\mathrm{kin}, \Theta_\mathrm{dyn}, \eta ) \nonumber \\
&& = {\cal N}( I_k(\Theta_\mathrm{dyn}) \cdot \mathrm{PED}[i](X_k(\Theta_\mathrm{kin}), Y_k(\Theta_\mathrm{kin}), \sigma_\mathrm{psf}), \sigma_S ) \nonumber
\end{eqnarray}
Here $I_k(\Theta_\mathrm{dyn})$ are the values of the selected parametrization of the light curve, and
\begin{eqnarray}
X_k(\Theta_\mathrm{kin}) &=& X_0 + U \cos(\Phi) (k-k_0), \label{eq:Xbayes}\\
Y_k(\Theta_\mathrm{kin}) &=& Y_0 + U \sin(\Phi) (k-k_0)\nonumber
\end{eqnarray}
are the coordinates of the model track point (for simplicity we limited ourselves to uniform motion, $A=0$), and the symbol $\mathrm{PED}[i]$ denotes the image distribution over the detector channels (from Pixel Energy Distribution). It depends both on the localization of the track point (the first two arguments) and on the~PSF. The PSF can be measured in a separate calibration experiment or modelled, for example, by an isotropic Gaussian distribution with a single unknown parameter~$\sigma_\mathrm{psf}$. The last variant is especially convenient because the PED is directly expressed through the error function~\texttt{erfc} implemented in the majority of mathematical libraries.

Thus, the nuisance parameters of our Bayesian model of track event are $\eta = \{ \sigma_S, \sigma_\mathrm{psf} \}$. For them, as well as for the parameters $\Theta$ that we are interested in within the reconstruction, a priori distributions must be defined (e.g., due to their positive definiteness, in the class of half-normals). The regularization priors on~$\Theta$ can be assigned on the basis of a previously performed image recognition procedure.

\section{Validation}
\label{sec:Validation}
The Bayesian track reconstruction methods were tested on both model and real events. 
Since~2022, our laboratory has deployed a PAIPS\footnote{PAIPS — Pulsating Aurora Imaging Photodetector System.} detector system in the Murmansk region, with photodetectors and electronics almost identical to JEM-EUSO ones. The~PAIPS modules consist of the same multi-anode PMTs ($8\times8$ pixels of 2.85~mm size) as in the~Mini-EUSO, with the same design of the elementary cell. The detectors, operating in monitoring mode with a temporal resolution of 1~ms, record a large number of track events: meteors, satellites and aircrafts. The~PAIPS data therefore provide a testbench for the development and validation of Bayesian reconstruction algorithms. The~PSF size at the Gaussian approximation is in the range of 0.5-1 mm. For more details on the~PAIPS detectors see~\cite{Klimov2022_PAIPS}. 

Reconstruction of simulated events that fall close to the centre of the field of view of a single~PMT, if the meteor generation model and the reconstruction model coincide, allows us to estimate the track speed with an accuracy above 1\% (with a typical signal-to-noise ratio for~PAIPS meteors). Replacing the model profile with an ``incorrect'' one leads to a systematic error in the track reconstruction, the magnitude of which depends on the total track length. Thus, when using the LIN+LIN profile (instead of the generative EXP+EXP) for short tracks (4-5 pixels long), the shift of the $\Phi$ angle was 1.5-2 degrees, and the error in speed estimation was up to 2.5\%, while for tracks of 6-8 pixels long the systematics is practically not noticeable. Similar results were obtained when the generated event had a non-Gaussian~PSF (Moffat distribution with different shape parameters \cite{Trujillo2001_MoffatPSF}) and was reconstructed with a Gaussian~PSF with a fitting value of~$\sigma_\mathrm{psf}$.

\begin{figure}[b]
\includegraphics[width=0.45\textwidth]{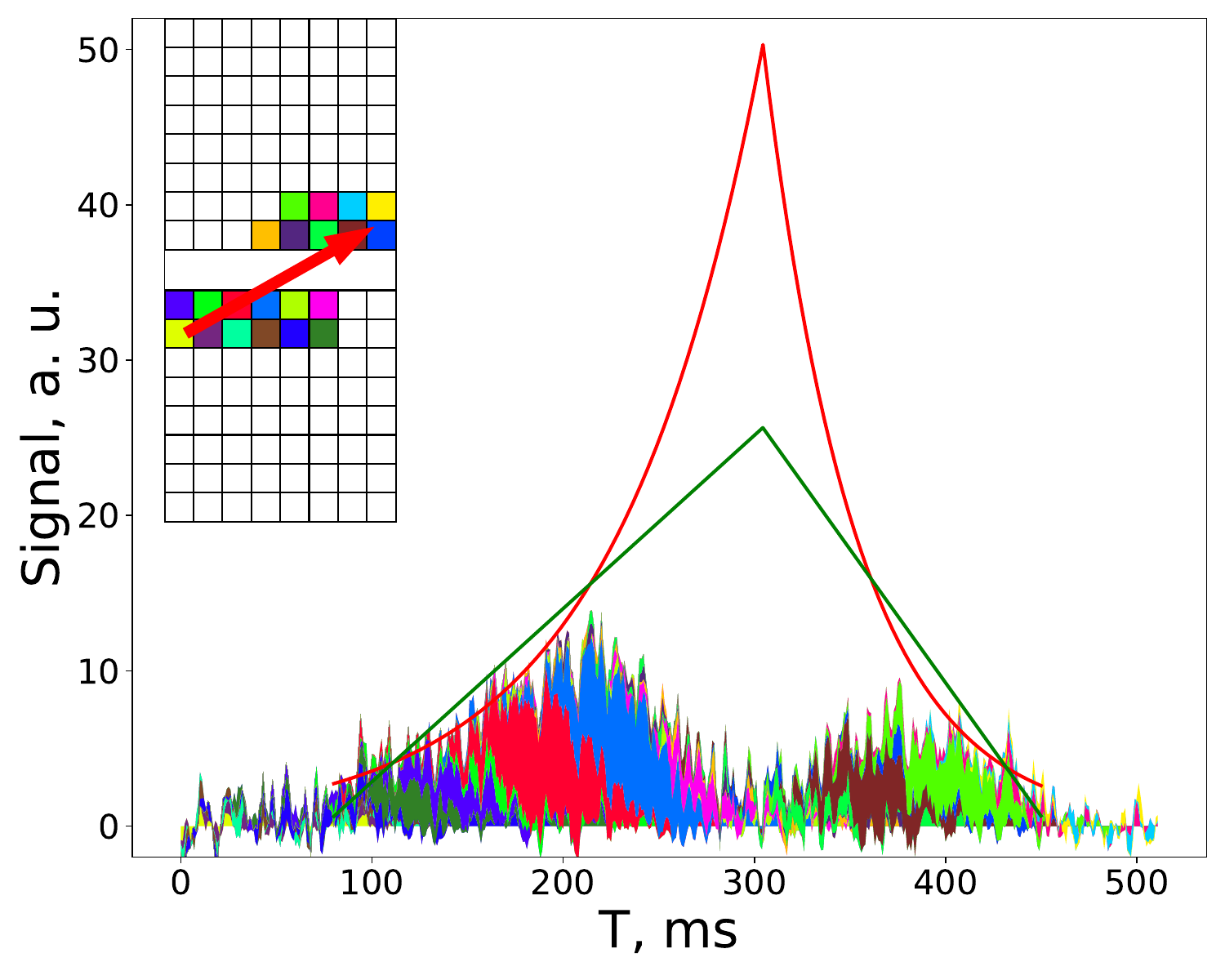}
\caption{\label{fig:1} Bayesian reconstruction of the simulated ``broken'' event with two profiles: EXP+EXP (red) and LIN+LIN (green). Active signals are shown in different colors as a stacked histogram (its envelope is the measured light curve). Inset: map of active channels and reconstructed track.}
\end{figure}

When the event is shifted to the edge of the field of view of a~PMT, and even when the track transits from one PMT to another (``broken'' event, the maximum of the signal falls into the dead zone and thus the average signal-to-noise ratio decreases by a factor of several), the Bayesian reconstruction is already noticeably superior to the barycentric one. In Figure~\ref{fig:1}, the ``broken'' model event is reconstructed using two Bayesian models - with the EXP+EXP profile (used in its generation) and with LIN+LIN one. It is interesting to note that, even in this case, the transfer of information by means of a parameterized profile that is only qualitatively similar to the true one allows to reconstruct the kinematic parameters of the event with high accuracy: the posterior distribution $p(U,\Phi | \mathrm{Data})$ is practically indistinguishable.

\begin{figure*}
\includegraphics[width=0.55\textwidth]{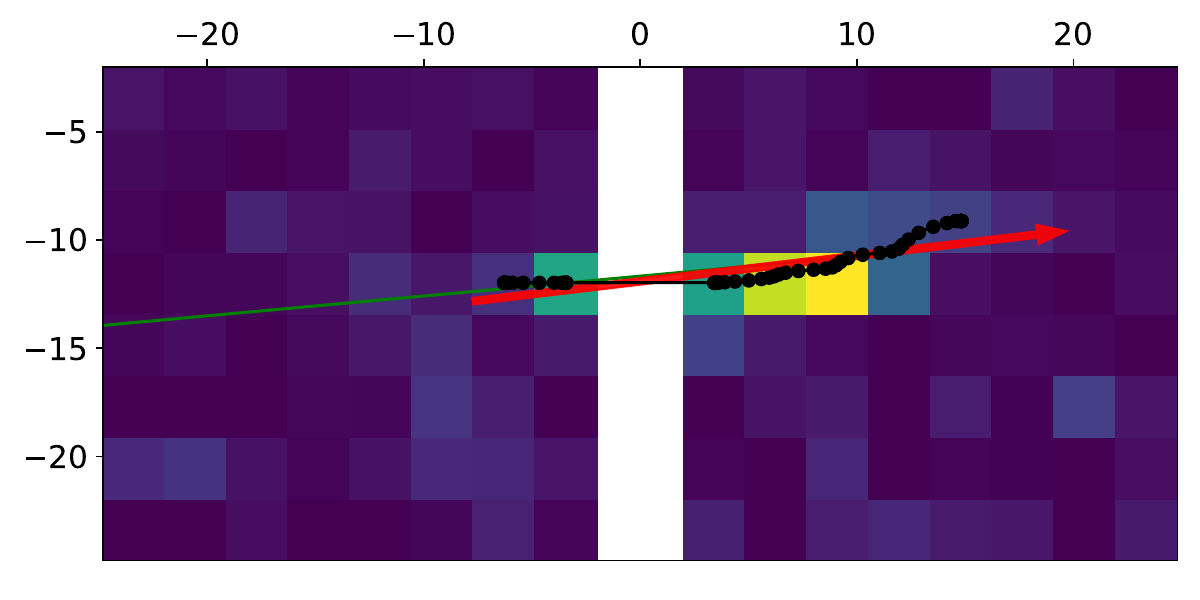}
\includegraphics[width=0.35\textwidth]{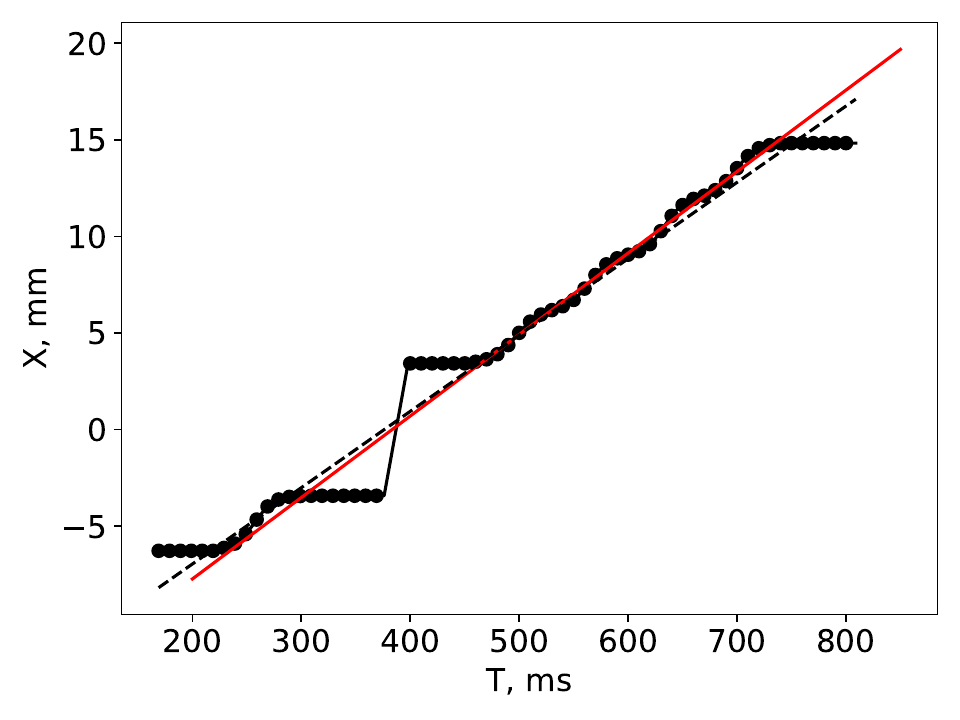}
\caption{\label{fig:2}Comparison of barycentric (black) and Bayesian (red) reconstructions of one of the Geminid-2022 meteors. Left - barycentric trajectories (dots) and reconstructed Bayesian track (arrow), green line - direction to the radiant. On the right, the corresponding $X(T)$ dependence (dashed line - least square estimate for the barycentres).}
\end{figure*}

Figure~\ref{fig:2} shows on the left a real event, meteor, recorded by one of the PAIPS detectors. The active channels have been distributed between two adjacent~PMTs, which greatly complicates the barycentric reconstruction. The barycentric trajectory is shown with black dots and the Bayesian reconstructed trajectory is shown with a red arrow. This event was recorded on the night of 12-13 December 2022, i.e. at the peak of the Geminid meteor shower \cite{Hajdukova2017_Geminids}. The green line shows the direction to the Geminid radiant (from the centre of the meteor): the difference to the reconstructed direction is only $\Delta\Phi = 1.5^\circ$ (and lies within the radiant dispersion).

The right panel shows both reconstructions, barycentric and Bayesian, using the (horizontal) $X$-coordinate of the~FS. As noted earlier, the barycentric dependence~$X_\mathrm{bar}(T)$ can differ significantly from the track-point motion at the beginning and end of the light curve and near the edge of the field of view (a similar systematics exists for the~$Y$-coordinate). The Bayesian dependence $X_\mathrm{bayes}(T)$ corresponds to~(\ref{eq:Xbayes}) with posterior averages for the $\Theta_\mathrm{kin}$ parameters. For this event, the differences in the recovered speed value were about~5\% ($U_\mathrm{bar} = 39.8$~mm/s vs. $U_\mathrm{bayes} = 42.0$~mm/s), the change in direction was 0.5$^\circ$ (closer to the Geminid radiant in the Bayesian case). 

\section{Conclusion}
\label{sec:Conclusion}
Probabilistic modelling can be very effective in situations where, for one reason or another, it is impractical to factor out any one type of information - geometric, kinematic and/or dynamic - from the data. This is particularly the case for ground-based and space-based multichannel detectors that build an image on a matrix of~PMTs. Dynamic images of transient phenomena have low spatial but high temporal resolution, and the focal plane of the photodetector contains numerous blind zones (structural gaps). The probabilistic model of the phenomenon includes both the parameters of the studied phenomenon and the measurement process (detector response function, distribution of measurement errors, etc.). The reconstruction of the phenomenon in this case consists in calculating the posterior distribution of the parameters (Bayesian inference) on the basis of the Bayes theorem. Averaging over nuisance parameters (marginalization rather than optimization) allows more reliable estimates to be obtained.

In this paper, probabilistic models are formulated and analyzed for two types of events: sub-millisecond ring-like glows in the upper atmosphere (ELVES) observed from Earth orbit, and track events (meteors, satellites, etc.) recorded by ground-based detectors. The Bayesian inference is implemented in probabilistic programming using the PyMC-5 library. The source code is open source and can be found at  \url{https://github.com/saraevrom/RecoV2}.

For shower meteors, the data can be supplemented by information on their direction and/or the characteristic speed (as an a priori distribution taking into account radiant dispersion).
This enables the reconstruction of the entire 3D-trajectory and and luminosity dynamics from  single detector data. At the same time, the Bayesian model can be easily generalized to stereo-data (signal registration by two or more detectors), which allows us to significantly reduce errors and reconstruct non-shower (sporadic) meteors. Another variant of the scheme modification allows to reconstruct EAS events; in this case the model parametrization should take into account that the light source itself moves at the speed of light. Such Bayesian reconstruction schemes are currently being tested on real PAIPS stereo-meteors and on simulated EAS events.



\section*{FUNDING}
This research was funded by Russian Science Foundation grant number 22-62-00010 (https://rscf.ru/project/22-62-00010/).

\section*{CONFLICT OF INTEREST}
The authors declare that they have no conflicts of interest.

\subsection*{Publisher's Note}
Pleiades Publishing remains neutral with regard to jurisdictional claims in published maps and institutional affiliations.




\begin{thebibliography}{0}%
\makeatletter
\providecommand \@ifxundefined [1]{%
 \@ifx{#1\undefined}
}%
\providecommand \@ifnum [1]{%
 \ifnum #1\expandafter \@firstoftwo
 \else \expandafter \@secondoftwo
 \fi
}%
\providecommand \@ifx [1]{%
 \ifx #1\expandafter \@firstoftwo
 \else \expandafter \@secondoftwo
 \fi
}%
\providecommand \natexlab [1]{#1}%
\providecommand \enquote  [1]{``#1''}%
\providecommand \bibnamefont  [1]{#1}%
\providecommand \bibfnamefont [1]{#1}%
\providecommand \citenamefont [1]{#1}%
\providecommand \href@noop [0]{\@secondoftwo}%
\providecommand \href [0]{\begingroup \@sanitize@url \@href}%
\providecommand \@href[1]{\@@startlink{#1}\@@href}%
\providecommand \@@href[1]{\endgroup#1\@@endlink}%
\providecommand \@sanitize@url [0]{\catcode `\\12\catcode `\$12\catcode `\&12\catcode `\#12\catcode `\^12\catcode `\_12\catcode `\%12\relax}%
\providecommand \@@startlink[1]{}%
\providecommand \@@endlink[0]{}%
\providecommand \url  [0]{\begingroup\@sanitize@url \@url }%
\providecommand \@url [1]{\endgroup\@href {#1}{\urlprefix }}%
\providecommand \urlprefix  [0]{URL }%
\providecommand \Eprint [0]{\href }%
\providecommand \doibase [0]{https://doi.org/}%
\providecommand \selectlanguage [0]{\@gobble}%
\providecommand \bibinfo  [0]{\@secondoftwo}%
\providecommand \bibfield  [0]{\@secondoftwo}%
\providecommand \translation [1]{[#1]}%
\providecommand \BibitemOpen [0]{}%
\providecommand \bibitemStop [0]{}%
\providecommand \bibitemNoStop [0]{.\EOS\space}%
\providecommand \EOS [0]{\spacefactor3000\relax}%
\providecommand \BibitemShut  [1]{\csname bibitem#1\endcsname}%
\let\auto@bib@innerbib\@empty
\end{thebibliography}%


\begin{thebibliography}{}

\bibitem{Coleman23_UHECR}
A. Coleman, J. Eser, E. Mayotte, et al.,
Astroparticle Physics. 149, 102819 (2023). https://doi.org/10.1016/j.astropartphys.2023.102819.


\bibitem{Barghini2022_TUSresults}
D. Barghini, M. Bertaina, A. Cellino, et al.,
Advances in Space Research. 70, 2734–2749 (2022). https://doi.org/10.1016/j.asr.2021.11.044. 

\bibitem{Adams2015_MiniEUSOScience}
J.H. Adams, S. Ahmad, J.N. Albert, et al.,
Exp. Astron. 40, 239-251 (2015)  https://doi.org/10.1007/s10686-014-9431-0.

\bibitem{Abdellaoui2018_EUSO-TA}
G. Abdellaoui, S. Abe, J.H. Adams, et al.,
Astropart. Phys., 102, 98 (2018). https://doi.org/10.1016/j.astropartphys.2018.05.007.

\bibitem{Adams2022_EUSO-Ball}
J.H. Adams, S. Ahmad, D. Allard, et al.,
Space Sci. Rev., 218, 3 (2022). https://doi.org/10.1007/s11214-022-00870-x.

\bibitem{Abdellaoui2024_EUSO-SPB1}
Abdellaoui, G., Abe, S., Adams, J. H., et al., Astropart. Phys., 154, 102891 (2024). https://doi.org/0.1016/j.astropartphys.2023.102891.

\bibitem{Klimov2022_KEUSO}
P. Klimov, M. Battisti, A. Belov, A., et al.,
Universe. 8, 88 (2022). https://doi.org/10.3390/universe8020088.

\bibitem{Olinto2021_POEMMA}
A.V. Olinto, J. Krizmanic, J.H. Adams, et al., 
J. Cosmol. Astropart. Phys., 2021, 007 (2021). https://doi.org/10.1088/1475-7516/2021/06/007.

\bibitem{Marcelli2024_MiniEUSOresults}
L. Marcelli. A. Belov, G. Garipov, et al.,
Instruments. 8(1), 6 (2024). https://doi.org/10.3390/instruments8010006.

\bibitem{Sivia2006_BI}
D. Sivia, J. Skilling, Data Analysis: A Bayesian Tutorial (Oxford science publications, 2006).

\bibitem{Martin2018_BI}
O. Martin, Bayesian Analysis with Python: Introduction to statistical modeling and probabilistic programming using PyMC3 and ArviZ (Packt Publishing, 2018).

\bibitem{Newsome2010_TLE}
R. T. Newsome,  U.S. Inan,
Journal of Geophysical Research: Space Physics. 115 (2010). https://doi.org/10.1029/2009JA014834.

\bibitem{Pasko2012_TLE}
V.P. Pasko, Y. Yair, C.L. Kuo,
Space Sci. Rev. 168, 475–516 (2012). https://doi.org/10.1007/s11214-011-9813-9.

\bibitem{Fukunishi1996_ELVES}
H. Fukunishi, Y. Takahashi, M. Kubota, et al.,
Geophysical Research Letters. 23, 2157-2160 (1996). https://doi.org/10.1029/96GL01979.

\bibitem{Barrington-Leigh1999_ELVES}
C.P. Barrington-Leigh, U.S. Inan,
Geophys. Res. Lett. 26, 683–686 (1999).  https://doi.org/10.1029/1999GL900059.

\bibitem{Klimov2019_TUSelves}
P. Klimov, B. Khrenov, M. Kaznacheeva, et al.,
Remote Sensing. 11, 2449 (2019). https://doi.org/10.3390/rs11202449.


\bibitem{Sharakin2024_RecoELVES}
S. Sharakin, D. Barghini, M. Battisti, et al.,
Cosmic Res. 62, 330–338 (2024). https://doi.org/10.1134/S0010952524600379.


\bibitem{Inan1991_Ionosph}
U.S. Inan, T.F. Bell, J.V. Rodriguez,
Geophysical Research Letters. 18, 705-708 (1991). https://doi.org/10.1029/91GL00364.

\bibitem{Loredo2024_BayesIntro}
T. Loredo, R.L. Wolpert,
Front. Astron. Space Sci. 11 (2024). https://doi.org/10.48550/arXiv.2406.18905.

\bibitem{Loredo2013_BAstrostatistics}
T.L. Loredo, Bayesian Astrostatistics: A Backward Look to the Future. In: Hilbe, J. (eds) Astrostatistical Challenges for the New Astronomy. (Springer Series in Astrostatistics, vol 1. Springer, New York, NY 2013). https://doi.org/10.1007/978-1-4614-3508-2\_2.

\bibitem{TUS-ELVES-1}
B.A. Khrenov, G.K. Garipov, M.Y. Zotov, et al., 
Cosmic Res 58, 317–329 (2020). https://doi.org/10.1134/S0010952520050056.

\bibitem{MiniEUSO-ELVES}
L. Marcelli, E. Arnone, M. Barghini, et al.,
Proceedings of 37th International Cosmic Ray Conference {\textemdash} PoS(ICRC2021). 395, 367 (2021).
https://doi.org/10.22323/1.395.0367.

\bibitem{Marshall2012_DoubleELVES}
R.A. Marshall, 
Journal of Geophysical Research: Space Physics. 117, A3 (2012). https://doi.org/10.1029/2011JA017408.

\bibitem{Barghini2024_MiniEUSOMET}
D. Barghini, M. Battisti, A. Belov, et al.,
Astronomy \& Astrophysics. 687, A304 (2024).  	https://doi.org/10.1051/0004-6361/202449236.

\bibitem{Ceplecha1987_MetRecoIP}
Z. Ceplecha,
Bulletin of the Astronomical Institutes of Czechoslovakia. 38, 222 (1987). https://ui.adsabs.harvard.edu/abs/1987BAICz..38..222C.

\bibitem{Borovicka1990_MetRecoLoS}
J. Borovicka,
Bulletin of the Astronomical Institutes of Czechoslovakia. 41, 391 (1990). https://ui.adsabs.harvard.edu/abs/1990BAICz..41..391B.

\bibitem{Gural2012_MetRecoMP}
P.S. Gural,
Meteoritics \& Planetary Science. 47, 1405-1418 (2012). https://doi.org/10.1111/j.1945-5100.2012.01402.x.

\bibitem{Vida2020_MetRecoMC}
D. Vida, P.S. Gural, P.G. Brown, et al.,
Monthly Notices of the Royal Astronomical Society. 491(2), 2688-2705 (2020). https://doi.org/10.1093/mnras/stz3160.

\bibitem{Pecina2009_MeteorsLC}
P. Pecina, P. Koten,
Astronomy \& Astrophysics. 499(1), 313-320 (2009).  https://doi.org/10.1051/0004-6361/200811503.

\bibitem{Lauer1999_PSF}
T. R. Lauer,
Publications of the Astronomical Society of the Pacific. 111(765), 1434 (1999).  https://doi.org/10.1086/316460. 

\bibitem{Klimov2022_PAIPS}
P. Klimov, S. Sharakin, A. Belov, et al.,
Atmosphere. 13(10), 1572 (2022). http://dx.doi.org/10.3390/atmos13101572.

\bibitem{Trujillo2001_MoffatPSF}
I. Trujillo, J. A. L. Aguerri, J. Cepa and C. M. Gutiérrez,
Mon. Not. R. Astron. Soc. 328, 977–985 (2001). https://doi.org/10.1046/j.1365-8711.2001.04937.x.

\bibitem{Hajdukova2017_Geminids}
M. Hajduková, P. Koten, L. Kornoš, J.Tóth,
Planetary and Space Science. 143, 89-98 (2017). https://doi.org/10.1016/j.pss.2017.01.004.



\end{thebibliography}
\end{document}